\documentclass[referee]{raa}    

\usepackage{amsmath,amssymb,amsxtra,amsfonts}
\usepackage{graphicx,times}

\usepackage{natbib}
\usepackage{txfonts}
\usepackage{footmisc}
\usepackage[USenglish]{babel}
\bibpunct{(}{)}{;}{a}{}{,}

\usepackage[colorlinks=true,linkcolor=blue,citecolor=blue]{hyperref}

\newcommand{\lcdm}{\mathrm{\Lambda CDM}}
\newcommand{\densm}{\Omega_{\mathrm{m}}}
\newcommand{\densl}{\Omega_{\mathrm{\Lambda}}}
\newcommand{\densk}{\Omega_{\mathrm{k}}}

\newcommand{\likelisym}{\mathcal{L}}
\newcommand{\G}{\mathcal{G}}
\newcommand{\xmodel}{\mathbf{x}}

\begin{document}

\title{Degeneracy and Discreteness in Cosmological Model Fitting}

\author{Huan-Yu Teng\inst{1}, Yuan Huang\inst{1}, Tong-Jie Zhang \inst{1,2,3}}

\institute{ 
Department of Astronomy, Beijing Normal University, Beijing 100875, P. R. China;{\it tjzhang@bnu.edu.cn}\\
\and
Departments of Physics and Astronomy, University of California, Berkeley, CA 94720, USA;\\
\and
Lawrence Berkeley National Laboratory, 1 Cyclotron Road, Berkeley, CA 94720, USA\\
}

\abstract{We explore the degeneracy and discreteness problems in the standard cosmological model ($\Lambda$CDM). We use the Observational Hubble Data (OHD) and the type Ia supernova (SNe Ia) data to study this issue. In order to describe the discreteness in fitting of data, we define a factor $\G$ to test the influence from each single data point and analyze the goodness of $\G$. Our results indicate that a higher absolute value of $\G$ shows a better capability of distinguishing models, which means the parameters are restricted into smaller confidence intervals with a larger figure of merit evaluation. Consequently, we claim that the factor $\G$ is an effective way in model differentiation when using different models to fit the observational data.
}

\authorrunning{Teng et al. }            
\titlerunning{Discreteness and degeneracy in cosmological fitting}  
\maketitle

\section{Introduction}
 
The \textit{PLANCK} \citep{2014A&A...571A..16P} satellite released its first results in 2013, which gave tighter constraints of cosmological parameter than before. Extensive observations have been made to constrain cosmological parameters including OHD \citep{2007MPLA...22...41Y,2014RAA....14.1221Z,2011ApJ...730...74M,2012JCAP...07..053M,2013PhLB..726...72F,2013ApJ...766L...7F,2015JCAP...02..025Y}, SNe Ia \citep{2012ApJ...746...85S,2003LNP...598..195P,1998AJ....116.1009R}, cosmic microwave background radiation (CMBR) \citep{2009ApJS..180..306D, 2011ApJS..192...18K, 2013ApJS..208...19H,2014A&A...571A..16P} and baryon acoustic oscillation (BAO) \citep{2005ApJ...633..560E, 2010MNRAS.401.2148P} . Qualitatively, the constraints imposed by more numerous observations can provide smaller confidence intervals of cosmological parameters. However, quantitative studies addressing how well the cosmological parameters are constrained if only limited datasets are available. In this paper, we present a new method of factor $\G$ to investigate this issue with OHD and SNe Ia Data and using the confidence interval and figure of merit (FoM) as inspection criteria.

\section{Methodology}
\label{sec:2}

\subsection{Standard Cosmological Model ($\Lambda$CDM)}
We examine a standard non-flat $\lcdm$ model with a curvature term, $\densk = 1 - \densm - \densl$, and without a radiation term \citep{2011ApJ...730...74M,2013PhLB..726...72F}. Specifically the Hubble parameter is given by
\begin{eqnarray}
    \label{eq:testmodel}
    H(z) &=& H_0 E(z; \densm, \densl, H_0) \nonumber 
    \\
    &=& H_0 \sqrt{\densm (1 + z)^3 + \densk (1 + z)^2 + \densl}.
\end{eqnarray}


The relationship between luminosity distance and redshift of SNe Ia is as below \citep{1998AJ....116.1009R,2011MNRAS.412.2685L} 
\begin{equation}
D_{L}=\frac{c(1+z)}{H_{0}} \mathrm{sinn}[ \sqrt{\vert \densk \vert} \int_0^{z'} \frac{1}{E(z; \densm, \densl, H_0)} dz']
\end{equation}

\begin{equation}
\mathrm{sinn}(x)=\left\{
\begin{array}{rcl}
\sinh x   &      & {\densk>0}\\
x         &      & {\densk=0}\\
\sin x    &      & {\densk<0}
\end{array} \right. \nonumber
\end{equation}

Noting that $\densm+\densl+\densk=1$, we have:
\begin{eqnarray}
\label{eq:3}
D_{L}= \frac{1+z}{\vert \sqrt{1- \densm - \densl} \vert} \mathrm{sinn} [  \sqrt{1-\Omega_{m}-\densl} \chi (z) ] \nonumber ,
\end{eqnarray}
\begin{eqnarray}
\chi (z)=\int_{0}^{z} \frac{d z'}{E(z; \densm, \densl,H_0)} .
\end{eqnarray}

The distance modulus is given by following an empirical equation \citep{1997ApJ...483..565P}:
\begin{eqnarray}
\label{eq:4}
\mu=5\log D_{L} - 5\log H + 52.384 .
\end{eqnarray}

Combining Equation (\ref{eq:3}) and (\ref{eq:4}), we obtain the relationship between distance modulus and redshift, which is dependent upon cosmological parameters.

Two datasets are utilized to constrain cosmological parameters, the existing 28 OHD \citep{2014RAA....14.1221Z,2005PhRvD..71l3001S,2010JCAP...02..008S,2012JCAP...07..053M,2013A&A...552A..96B,2012MNRAS.425..405B,2013MNRAS.435..255C} and the SNe Ia data provided by Supernova Cosmology Project (SCP) \citep{2012ApJ...746...85S}, which contains 580 type Ia supernovae with redshifts, distance modulus and errors. 

\subsection{Degeneracy and Discreteness}
\begin{figure}
\centering
\includegraphics[width=.80\textwidth]{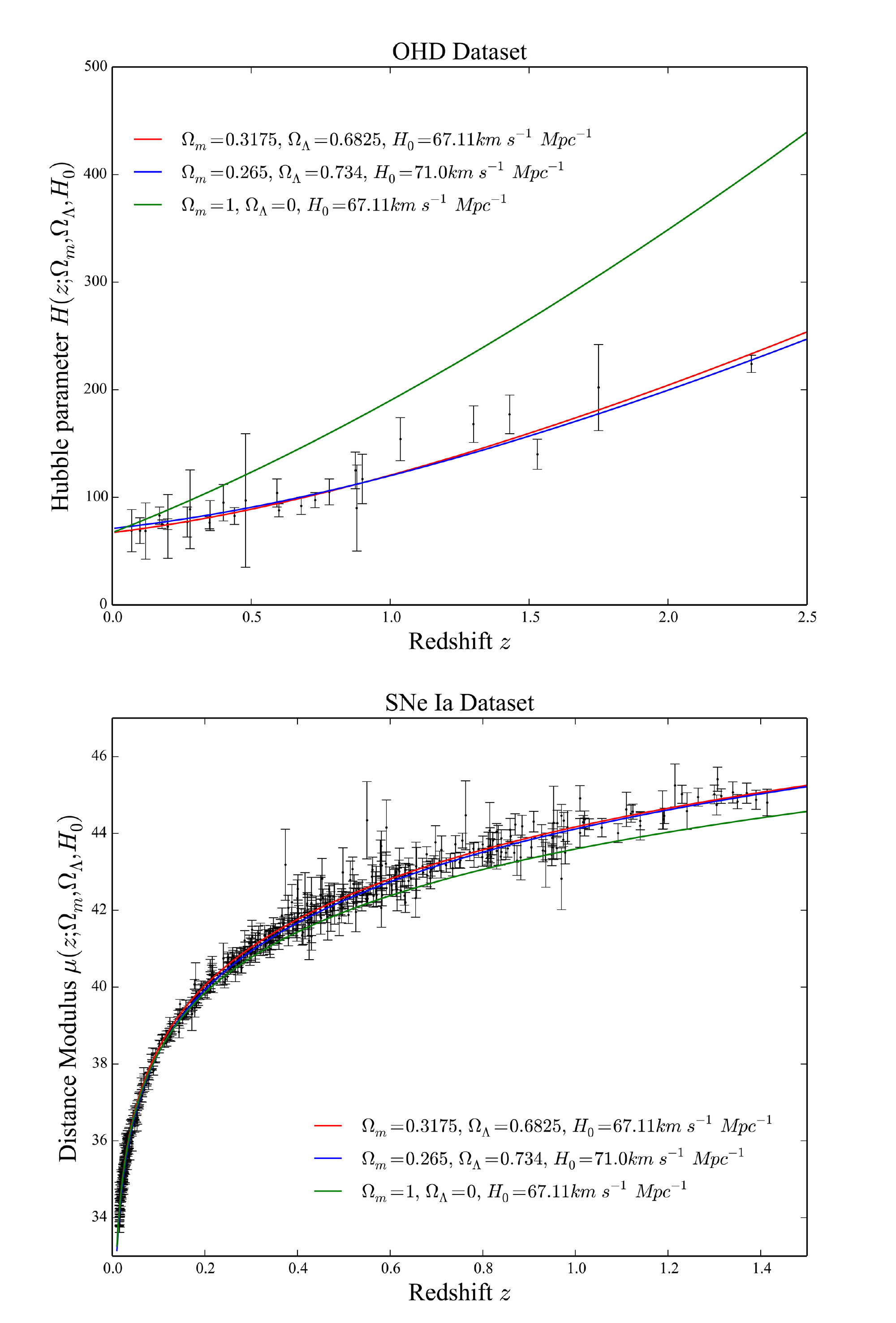}
\caption{Theoretical and observational Hubble parameter and the distance modulus of SNe Ia. 
The upper sub-figure shows the OHD dataset with $1\sigma$ confidence interval and the theoretical $H(z; \densm, \densl,H_0)$ value of different models. 
The lower one indicates the SNe Ia datasets with $1\sigma$ confidence interval and the theoretical curves predicted by different models.
The red, blue and green curves represent the model predictions \textit{PLANCK} \citep{2014A&A...571A..16P}, WMAP \citep{2011ApJS..192...18K}, and the model of full matter, respectively. Black dots and error bars indicates observational data.
}
\label{fig:1}
\end{figure}

\begin{figure}
\centering
\includegraphics[width=.80\textwidth]{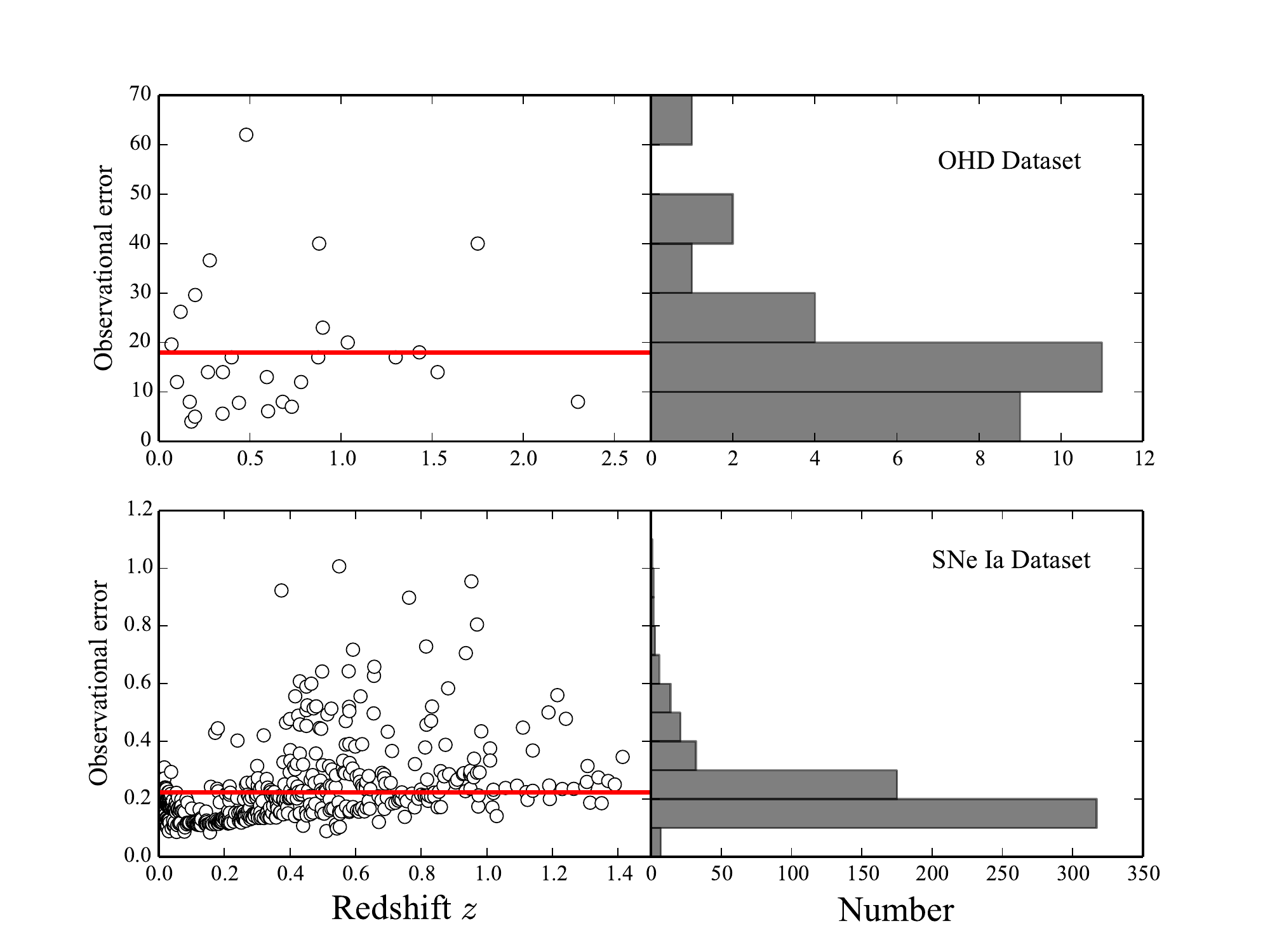}
\caption{The observational error of SNe Ia dataset given by SCP. The top sub-figures illustrate the OHD dataset and the bottom sub-figures iluustrate the SNe Ia datasets. Besides, the left sub-figures show the two dimensional diagrams between observational error and redshift, and the red lines are the average values of all errors. The right sub-figure shows the histogram of the distribution of observational errors (OHD in 10 bins and SNe Ia in 12 bins).}
\label{fig:2}
\end{figure}

How the Hubble parameter and the distance modulus depends on redshift are shown in Figure \ref{fig:1}. An inspection of Figure \ref{fig:1} suggests that in low-redshift regions, different models predict very similar distance modulus, i.e., they are degenerate. Therefore, the OHD and SNe Ia data in low-redshift regions cannot be used to distinguish these models. 
Here we use Figure \ref{fig:2} to show the observational error of OHD and SNe Ia dataset since error bars is not clearly shown in Figure \ref{fig:1}. From figure \ref{fig:2}, we find that the values observational errors are basically at the level of the red line, and in the region of OHD is from $0$ to $30$ meanwhile the region of SNe Ia is from $0.1$ to $0.3$. Besides, the Figure \ref{fig:2} does not shows a obvious relationship between observational error and redshift.

\label{subsec:DD}

Based upon the likelihood function of single point, we can study the relationship between the goodness-of-fit and the final fitting results for a given model. With a set of data for fitting, the probability densities of each point show little difference in parameter space, suggesting that the probabilities of all parameters are approximately the same. We cannot distinguish the best fitting points, and meanwhile the confidence intervals are relatively large. In order to study the quality of the single points, we make a perturbation of one parameter to investigate how the likelihood of the single point varies. This can be seen as finding the absolute value of derivative of the likelihood. Likelihood depends on parameters (We give the equations in section \ref{subsec:G}), which indicates that we can calculate the derivative of likelihood with respected parameters.

For the OHD dataset, we calculate the partial derivative of $H(z;\densm,\densl, H_0)$ with respect to different parameters. In this paper, we do not consider the goodness of confidence interval of $H_0$.

\begin{eqnarray}
\frac{\partial H}{\partial \densm}=\dfrac{H_{0}z(1+z)^{2}}{2 \sqrt{\densm (1+z)^{3}+ \densl + (1-\densm-\densl)(1+z)^{2}}}
\end{eqnarray}

\begin{eqnarray}
\frac{\partial H}{\partial \densl}=\dfrac{-H_{0}z(z+2)}{2 \sqrt{\densm (1+z)^{3}+ \densl + (1-\densm-\densl)(1+z)^{2}}}
\end{eqnarray}

For the SNe Ia dataset, we calculate the partial derivative of distance modulus with respected order parameters: $\mu=\mu(z; \densm, \densl,H_0)$:

\begin{center}
\begin{eqnarray}
\frac{\partial \mu }{\partial \densm}=\frac{5}{\ln 10} \frac{1}{D_{L}}(\frac{D_{L}}{2\densk} + (1+z) \mathrm{cosn}(\vert \sqrt{\densk} \chi(z) \vert) \\ 
(-\frac{1}{2}\frac{\chi(z)}{2\densk}
- \int_{0}^{z} \frac{dz}{2E^{3}} z(z+1)^{2}))  \nonumber 
\end{eqnarray}
\end{center}

\begin{eqnarray}
\frac{\partial \mu}{\partial \densl}=\frac{5}{\ln 10} \frac{1}{D_{L}}(\frac{D_{L}}{2\densk}+(1+z) \mathrm{cosn}(\vert \sqrt{\densk} \chi(z) \vert) \\  
(-\frac{1}{2}\frac{\chi(z)}{2\densk} - \int_{0}^{z} \frac{dz}{2E^{3}} z(z+2)))   \nonumber 
\end{eqnarray}

Noticing that in the above four functions $\densm+\densl+\densk=1$ is assumed still, and the definition of $\mathrm{cosn}(x)$ is similar to $\mathrm{sinn}(x)$.

\begin{equation}
\mathrm{cosn}(x)=\left\{
\begin{array}{rcl}
\cosh x   &      & {\densk>0}\\
x         &      & {\densk=0}\\
\cos x    &      & {\densk<0}
\end{array} \right. \nonumber
\end{equation}

\subsection{Definition of Factor $\G$}
\label{subsec:G}

In the next following equations, we introduce $\theta$ to represent $\densm$ or $\densl$, and $\xmodel$ represent the observational variables, $\mu$ or $H$. For the sake of simplicity, we use a subscript $th$ to denote theoretical values, $ob$ to denote observation values, i.e $\xmodel_{th}$ and $\xmodel_{ob}$. $\likelisym$ is the symbol of likelihood, i.e.,

\begin{eqnarray}
\likelisym_{i}(z;\theta \vert \xmodel) = \exp{[-\frac{(\xmodel_{ob,i}-\xmodel_{th,i})^{2}}{ 2 \sigma_{i}^{2}}]}
\end{eqnarray}

\begin{eqnarray}
\frac{\partial}{\partial \theta} (-\ln \likelisym_{i}) = \frac{(\xmodel_{ob,i}-\xmodel_{th,i})}{\sigma_{i}^{2}} \frac{\partial \xmodel_{th,i}}{\partial \theta}
\end{eqnarray}

The posterior of a model is proportional to the product of the likelihood of each point:

\begin{eqnarray}
\frac{\partial}{\partial \theta} (-\ln \likelisym) = \frac{\partial }{\partial \theta} (-\ln \prod_{i=1}^{n} \likelisym_{i})
\end{eqnarray}

\begin{eqnarray}
\frac{\partial}{\partial \theta} (-\ln \likelisym) = 
\frac{\partial \xmodel_{th,n}}{\partial \theta}\frac{(\xmodel_{ob,n}-\xmodel_{th,n})}{\sigma_{n}^{2}} \\
 + \sum_{i=1}^{n-1}(\frac{\partial \xmodel_{th,i} }{\partial \theta}\frac{(\xmodel_{ob,i}-\xmodel_{th,i})}{\sigma_{i}^{2}})  \nonumber
\end{eqnarray}

Since the observational error is stochastic, randomly up and down to the difference between $\xmodel_{ob,i}$ and $\xmodel_{th,i}$. Moreover, only the gradient effect is focused on in our study. Hence we set $\xmodel_{ob,i}-\xmodel_{th,i} = \sigma = \bar{\sigma}$, where $\bar{\sigma}$ is the average of observational errors of all data.
Under the same condition, we can define the discreteness factor $\G$, that is proportional to $\frac{1}{\sigma}\frac{\partial \xmodel_{th}}{\partial\theta}$:

\begin{eqnarray}
\G(z;\theta \vert \xmodel) = \frac{1}{\sigma}\frac{\partial \xmodel_{th}}{\partial\theta}
\label{eq:G}
\end{eqnarray}

The factor $\G$ depends on the redshifts and the confidence intervals of the data. Small confidence intervals are required to distinguish models if the models tend to be degenerate. In the parameter regions where the models appear to be discrete, we do not need small confidence intervals. The $\G$ value can be used to quantitatively measure the discreteness of data points with errors at different redshifts.

For a fitting process, if a new observational data point is added, it will take such an effect to the result:

\begin{eqnarray}
\G_{n} = \frac{\partial }{\partial \theta}(-\ln \likelisym_{n})
\end{eqnarray}

\begin{eqnarray}
\frac{\partial}{\partial \theta} (-\ln \likelisym) = 
\G_{n} +  \sum_{i=1}^{n-1}[\frac{\partial \xmodel_{th,i} }{\partial \theta}\frac{(\xmodel_{ob,i}-\xmodel_{th,i})}{\sigma_{i}^{2}}] .
\end{eqnarray}

Here we use $-\ln \likelisym(z, \theta)$ instead of $\likelisym(z, \theta)$ to simplify formulas. It is not so important to find an explicit equation of likelihood. Now we have a criterion to measure the discreteness of different data points caused by the model itself. 
In next section, we will apply our method to the existing 28 OHD and 580 SNe Ia data given by SCP to examine the power of the factor $\G$.   

\begin{figure}
\centering
\includegraphics[width=.80\textwidth]{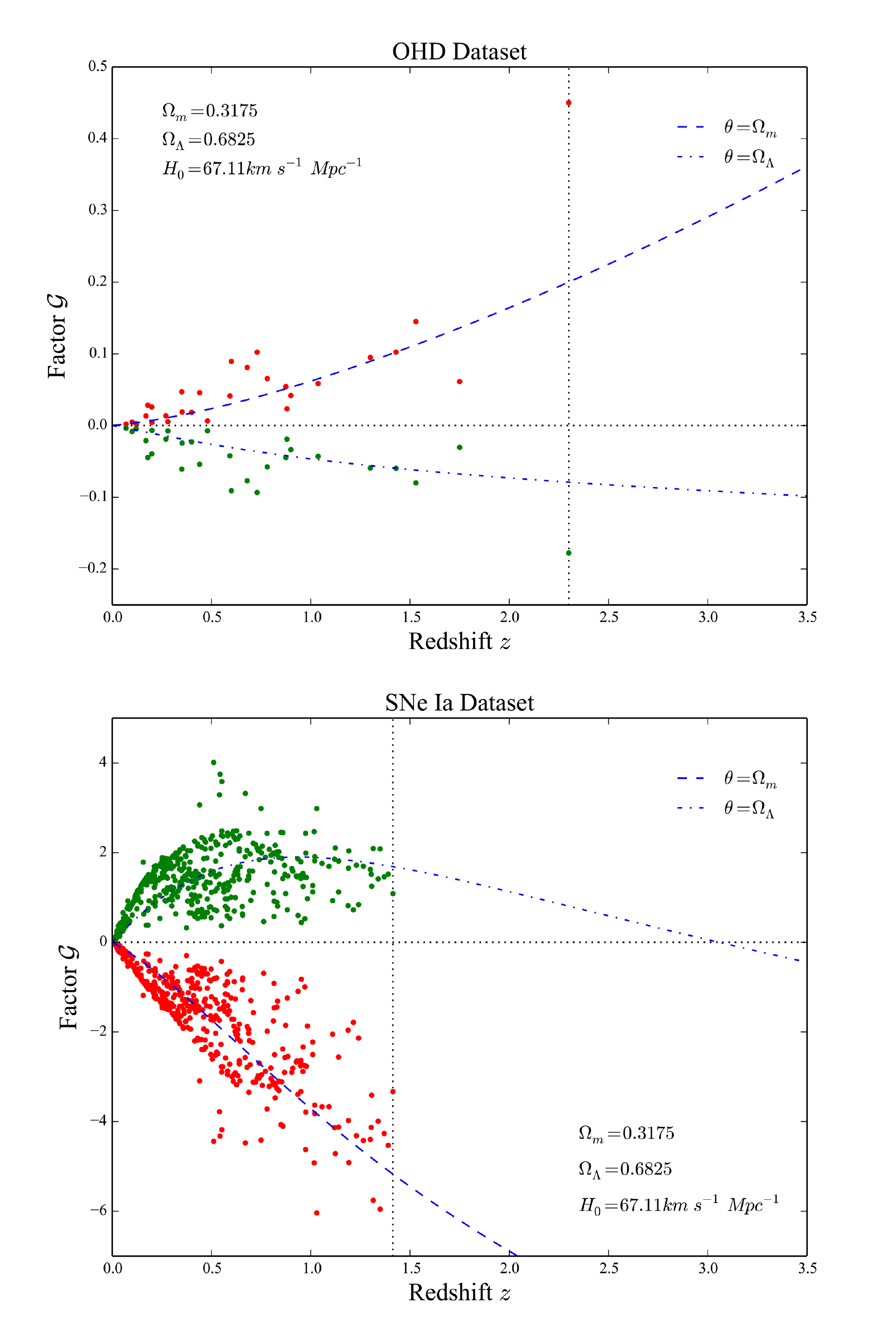}
\caption{ Theoretical and observational factor $\G$ of OHD set and SNe Ia dataset.
In these two sub-figures, blue dashed and dotted-dashed curves represent the theoretically predictions. We take OHD set and SNe Ia dataset into Equation \ref{eq:G} to calculate factor $G(z;\densm \vert H)$, $\G(z;\densm \vert \mu)$, $G(z;\densl \vert H)$ and $\G(z;\densl \vert \mu)$ of each point, then we plot these $\G$s as dots in the figure. The vertical black dotted line indicates the point of highest redshift and the horizontal black dotted line denotes $\G=0$.
}
\label{fig:3}
\end{figure}

\section{Analysis and Results}
\label{sec:3}

One of the results that we focus on is the relationship between the redshift $z$ and the factor $\G$. 
From the definition of the factor $\G$, we find it is also related to the parameters we select, which means, for different parameters at the same redshift, the goodness of discreteness of the data points is different. 

Nowadays constraints on cosmological parameters give the value of $\densm \simeq 0.32$ and $\densl \simeq 0.68$ \citep{2014A&A...571A..16P}, i.e. \textit{PLANCK}.
In our experiment, We find that these values are hardly affected by adding more datasets. After parameters confirmed, the factor $\G$ will be a function only relying on the redshift and observational error.

\begin{table}
\label{tab:1}
\tabcolsep 0pt
\vspace*{-10pt}
	\begin{center}
		\def\temptablewidth{0.8\textwidth}
	{\rule{\temptablewidth}{1pt}}
\begin{tabular*}{\temptablewidth}{@{\extracolsep{\fill}}ccccccc}

    Dataset &$\theta$  &  Lev.     &  Dat.   &  $\densm$               &  $\densl$               & $FoM_{m \Lambda}$ \tabularnewline
    \hline
    SNe Ia  &   &  all      &  $580$  &  \emph{0.273$\pm$0.070} &  \emph{0.712$\pm$0.117} & 387.18 \tabularnewline
    SNe Ia  &$\densm$  &  high     &  $500$  &  \emph{0.287$\pm$0.071} &  \emph{0.751$\pm$0.126} & 353.54 \tabularnewline
    SNe Ia  &$\densm$  &  high     &  $400$  &  \emph{0.255$\pm$0.086} &  \emph{0.657$\pm$0.187} & 206.13 \tabularnewline
    SNe Ia  &$\densm$  &  mid      &  $500$  &  \emph{0.320$\pm$0.120} &  \emph{0.788$\pm$0.168} & 200.93 \tabularnewline
    SNe Ia  &$\densm$  &  mid      &  $400$  &  \emph{0.279$\pm$0.179} &  \emph{0.784$\pm$0.223} & 109.64 \tabularnewline
    SNe Ia  &$\densm$  &  low      &  $500$  &  \emph{0.222$\pm$0.152} &  \emph{0.672$\pm$0.187} & 142.60 \tabularnewline
    SNe Ia  &$\densm$  &  low      &  $400$  & -1.097$\pm$0.357 &  0.342$\pm$0.353  &  \tabularnewline
    SNe Ia  &$\densl$  &  high     &  $500$  &  \emph{0.281$\pm$0.072} &  \emph{0.739$\pm$0.127} & 353.54 \tabularnewline
    SNe Ia  &$\densl$  &  high     &  $400$  &  \emph{0.270$\pm$0.081} &  \emph{0.706$\pm$0.169} & 238.91 \tabularnewline
    SNe Ia  &$\densl$  &  mid      &  $500$  &  \emph{0.291$\pm$0.079} &  \emph{0.778$\pm$0.132} & 303.60 \tabularnewline
    SNe Ia  &$\densl$  &  mid      &  $400$  &  \emph{0.320$\pm$0.089} &  \emph{0.810$\pm$0.147} & 210.99 \tabularnewline
    SNe Ia  &$\densl$  &  low      &  $500$  &  \emph{0.307$\pm$0.087} &  \emph{0.743$\pm$0.137} & 244.70 \tabularnewline
    SNe Ia  &$\densl$  &  low      &  $400$  &  \emph{0.245$\pm$0.139} &  \emph{0.780$\pm$0.189} & 126.50 \tabularnewline
    OHD     &  &  all      &  $28 $  &  \textbf{\emph{0.279$\pm$0.078}} & \textbf{\emph{0.637$\pm$0.260}}  & \textbf{\emph{128.63}}\tabularnewline
    OHD     &$\densm$  &  high     &  $22 $  &  \emph{0.280$\pm$0.082} &  \emph{0.643$\pm$0.279} & 125.11\tabularnewline
    OHD     &$\densm$  &  high     &  $16 $  &  \emph{0.271$\pm$0.097} &  \emph{0.438$\pm$0.416} & 78.18\tabularnewline
    OHD     &$\densm$  &  mid      &  $22 $  &  0.870$\pm$0.220 &  1.480$\pm$0.369  &  \tabularnewline
    OHD     &$\densm$  &  mid      &  $16 $  &  0.779$\pm$0.409 &  1.230$\pm$0.661  &  \tabularnewline
    OHD     &$\densm$  &  low      &  $22 $  &  0.740$\pm$0.391 &  1.220$\pm$0.610  &  \tabularnewline
    OHD     &$\densm$  &  low      &  $16 $  &           &         &   \tabularnewline
    OHD     &$\densl$  &  high     &  $22 $  &  \textbf{\emph{0.293$\pm$0.072}} & \textbf{\emph{0.766$\pm$0.235}}  & \textbf{\emph{150.92}} \tabularnewline
    OHD     &$\densl$  &  high     &  $16 $  & -0.202$\pm$1.569 & -0.181$\pm$2.072 & 100.72 \tabularnewline
    OHD     &$\densl$  &  mid      &  $22 $  &  \emph{0.258$\pm$0.091} &  0.497$\pm$0.422 & 79.31 \tabularnewline
    OHD     &$\densl$  &  mid      &  $16 $  &           &          &  \tabularnewline
    OHD     &$\densl$  &  low      &  $22 $  &           &          &  \tabularnewline
    OHD     &$\densl$  &  low      &  $16 $  &           &           & \tabularnewline

\end{tabular*}
{\rule{\temptablewidth}{1pt}}
\end{center}
\vspace*{-18pt}
\caption{The fitting results by reordering all data depending on $\G$. In column $\densm$ and $\densl$: 
(1) Emphasized text means reasonable in $\Lambda$CDM model.
(2) Normal text means unreasonable in $\Lambda$CDM model.
(3) Blanks of ($\densm, densl$) means no constraint results due to non-convergence.
(4) Results of ($\densm, densl$) in bold text and result in the first row of the table are shown in figure \ref{fig:4}. 
}

\end{table}

Figure \ref{fig:3} shows how factor $\G$ changes with redshift $z$. Here, for the theoretical value of factor $\G$, we assume that the all standard deviations errors are equal to the average standard deviation error. 
For observational value of factor $\G$, we take observational standard errors. Since the key point is the degree that likelihood changes with the cosmological parameter, it is sufficient to focus on the absolute value $\vert \G \vert$.

From Figure \ref{fig:3}, we find that $\vert G(z;\densm  \vert \xmodel )\vert$ and $\vert G(z;\densl  \vert H)\vert$ monotonically increase with redshift, while $\vert G(z;\densl  \vert \mu)\vert$ does not, which indicates that for the constraint on $\densm$, high redshift data show the obvious advantage. However, this is not the case for $\densl$, high redshift datasets of SNe Ia do not show this kind of advantage. Considering the true standard deviation error of each data point, Figure \ref{fig:3} also indicates the true value of discreteness. Both true and theoretical deviations error indicate the same tendency, but due to the different errors of each data point, there is a fluctuation in goodness of discreteness. Consequently, although some data points are at higher redshifts, the factor $\G$ of these data points may be smaller than $\G$s of lower redshift data points.

Next, we reorder data depending on the absolute value of factor $\G$ of $\densm$ and $\densl$ for both OHD and SNe Ia data. To make comparison, we take different factor $\G$ of different parameters. For OHD, we take 28, 22, and 16 out of 28 data and while for SNe Ia, we take 580, 500 and 400 out of 580 data. When not all data are selected, we linearly divide data into three levels, the high, the mid and the low $\G$s of all data. Then we employ MCMC method to resample the best-fitting point and explore the changes of the best-fitting point and confidence interval. We use the publicly available code \texttt{PyMC}
(https://github.com/pymc-devs/pymc)
to perform a full MCMC analysis. The results are listed in Table 1. The results close to \textit{PLANCK} \citep{2014A&A...571A..16P} and WMAP \citep{2011ApJS..192...18K} can be considered reasonable and the reasonable results are also close to results using SNe Ia data and OHD and given by \cite{2011ApJS..192...18K,2011ApJ...730...74M,2012JCAP...07..053M,2012JCAP...11..018W,2013PhLB..726...72F,2014RAA....14.1221Z}.

Table 1 
indicates that if we select same level of data, for both $\densl$ and $\densm$, the confidence intervals are generally increasing with the quantities of datasets decreasing, suggesting that our factor $\G$ is effective in distinguishing the data with discreteness. However, we find that in some cases the Markov Chains of this group of data does not converge, or the fitting yields abnormal results.

Theories of probability and statistics indicate that that the goodness-of-fit increases with the amount of observational data. However, in our experiment, we find that the confidence intervals may become smaller when we remove data of lower factor $\G$. There are two ways to datasets' ability to tigthen the constrains, directly checking the standard deviation error of the constrained parameter to examine the constraint on the specified parameter and establishing a quantified figure of merit (FoM) to examine the comprehensive effect on both parameters. The FoM can be defined as long as it reasonably rewards a tight fit while punishing a loose one. We apply the definition of $0.95$ confidence region in a parameter space \citep{2006astro.ph..9591A}, which can be calculated by:

\begin{eqnarray}
FoM_{xy} = \frac{\pi}{A} = \frac{1}{ \sigma(\theta_{x}) \sigma(\theta_{y}) \sqrt{1-\rho_{xy}}}
\end{eqnarray} 

Here $\rho_{xy}$ is correlation coefficient between $\theta_{x}$ and $\theta_{y}$, with the relation to covariance matrix $ \mathbf{C}_{x y}=\sigma(\theta_{x}) \sigma(\theta_{y}) \rho_{xy}$. The larger FoM is, the better constraint we get.

Seeing that in the OHD dataset, we have $\sigma=0.235$ of 22 higher part, smaller than $\sigma=0.260$, if we set $\densl$ as the parameter of $\G$, and in this case, we consider that OHD at $z=0.48$, $z=0.88$ and $z=1.75$ did not take positive effect altogether in fitting. Therefore we may pick all these negative points out theoretically. Besides, we notice that the $FoM_{m \Lambda}$ of 22 higher part is larger than the one of all data, which means the declination of the points even improve the total constraint quality.

Moreover, we should notice that the standard deviation error of $\densm$ and $\densl$ contains a correlation due to the same $\sigma^{-1}$. Here, a point of larger discreteness of $\densm$ may show a relatively larger discreteness of $\densl$.

\begin{figure}
\centering
\includegraphics[width=.80\textwidth]{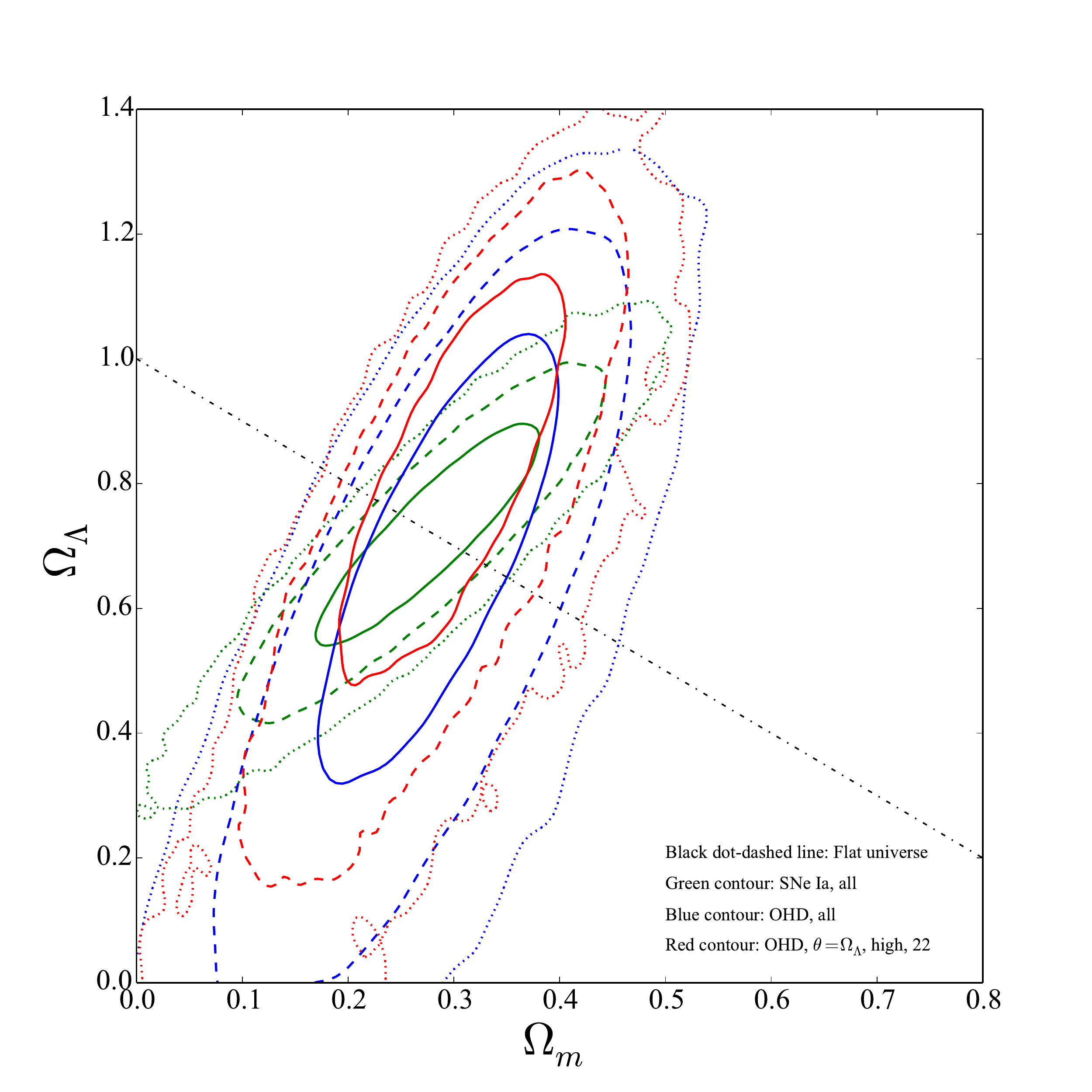}
\caption{Typical confidence regions in the $(\densl,\densm)$ parameter subspace. 
The solid, dashed and dotted contours respectively correspond to $68.3 \% $, $95.4\% $ and $99.7\% $ intervals. The black dot-dashed line indicates a flat universe. The $\densl$ interval of red contour is slightly smaller than blue contour which is marked bold in Table 1.}
\label{fig:4}
\end{figure}

We should mention that the $\G$ factor applies for one parameter in one function, and the effect brought by data removal has been ignored. It can be considered that the factor $\G$ we defined reflects the quality of the observational data. We remove some data in order to discuss the property of the factor $\G$ and find data that do not have or positive effect on the fitted cosmological model. Furthermore, we deal with these removed points in accordance with specific conditions.

\section{Conclusion and Discussion}
\label{sec:4}
In this paper, we develop a new method to study the degeneracy and discreteness in cosmological model. We define a criterion called discreteness factor $\G$ that relates the modeling functions, likelihood and undetermined parameter. The definition of the factor $\G$ is independent of any specific modeling functions, hence it can be generalized to an arbitrary modeling process. We start from non-flat $\Lambda$CDM model based on the existing 28 observational Hubble data (OHD)\citep{2014RAA....14.1221Z,2005PhRvD..71l3001S,2010JCAP...02..008S,2012JCAP...07..053M,2013A&A...552A..96B,2012MNRAS.425..405B,2013MNRAS.435..255C} and 580 type Ia supernova (SNe Ia) data released by Supernova Cosmology Project (SCP)\citep{2012ApJ...746...85S}. The functions indicate that theoretically factor $\G$ of $\densm$ increases with redshift, however, due to the different observational standard deviations errors for all data, true value of $\G$s only shows the trend, especially for the OHD dataset. 

We compute the factor $\G$ for $\densm$ and $\densl$ and reorder the data utilizing true value of $\G$s in both OHD and SNe Ia datasets. We generate Markov Chain Monte Carlo (MCMC) to find the best-fitting points and their confidence intervals. The fitting results demonstrate that the $\G$ takes effect on fitting, and higher absolute value of $\G$ gives a stronger constraint and a larger FoM evaluation. Besides, data of lower $\G$ values may provide not only larger confidence interval but also unreasonable best-fitting point. But the factor $\G$ displays its limitation in some aspect. According to theory of statistics and probability, the confidence interval explains that some data are removed usually. However, as if the effect of $\G$ is strong enough, it will cover the intrinsic properties of statistics and probability. Once we find the intervals decrease or FoM increases with fewer data of lower $\G$ value, we may find observational data hardly taking positive effects in constraint. 

\section*{Acknowledgments}
We are grateful to Jin Wu, Yong Zhang, Hao-Feng Qin and Ze-Long Yi for the improvement of the paper. Tong-Jie Zhang thank Prof. Martin White for his hospitality during his visit at Departments of Physics and Astronomy, University of California, Berkeley and Lawrence Berkeley National Laboratory. This work was supported by the National Science Foundation of China (Grants No. 11173006), the Ministry of Science and Technology National Basic Science program (project 973) under grant No. 2012CB821804.

\clearpage

\bibliography{ms}

\bibliographystyle{raa}

\end{document}